\documentclass[a4paper, 11pt]{article}
\usepackage[utf8]{inputenc}
\usepackage{indentfirst}
\usepackage{geometry}
\usepackage{textcomp}
\usepackage{amsmath}
\usepackage{amssymb}
\usepackage{graphicx}
\usepackage{color,xcolor}
\usepackage{mathrsfs}
\usepackage{caption}
\usepackage{dsfont}
\usepackage[colorlinks=true, allcolors=blue]{hyperref}
\usepackage{tcolorbox}
\usepackage{tikz}
\usepackage{tikz-cd}
\usepackage{physics}
\usepackage{diagbox}
\usepackage{listings}
\usepackage{appendix}
\usepackage{array}
\usepackage{subfig}

\usepackage{cite}

\catcode`\>=\active \def>{
\fontencoding{T1}\selectfont\symbol{62}\fontencoding{\encodingdefault}}
\newcommand{\nobracket}{}
\newcommand{\tmop}[1]{\ensuremath{\operatorname{#1}}}

\usepackage{amsmath}
\numberwithin{equation}{section}
\usepackage[normalem]{ulem}
\usepackage{color}
\definecolor{darkgreen}{RGB}{40,150,60}

\def\D{\Delta}

\def\m{\mu}

\def \hs{\hspace}

\def\be{\begin{equation}}
\def\ee{\end{equation}}
\def\bag{\begin{aligned}}
	\def\eag{\end{aligned}}
\def\bea{\begin{eqnarray}}
\def\eea{\end{eqnarray}}
\def\ba{\begin{array}}
	\def\ea{\end{array}}

\title{Bulk reconstruction in flat holography}
\author{Bin Chen$^{1,2,3}$, and Zezhou Hu$^1$}
\date{\today}

\begin{document}

\maketitle
\begin{center}
	{\it
		$^{1}$School of Physics, Peking University, No.5 Yiheyuan Rd, Beijing 100871, P.~R.~China\\
		\vspace{2mm}
		$^{2}$Collaborative Innovation Center of Quantum Matter, No.5 Yiheyuan Rd, Beijing 100871, P.~R.~China\\
		$^{3}$Center for High Energy Physics, Peking University, No.5 Yiheyuan Rd, Beijing 100871, P.~R.~China\\
	}
	\vspace{10mm}
\end{center}

\begin{abstract}
    \vspace{5mm}
   In this note, we discuss the bulk reconstruction of massless free fields in flat space from the highest-weight representation of boundary Carrollian conformal field theory (CCFT). We expand the bulk field as a sum of infinite descendants of a primary state defined in the boundary CCFT, and discuss the Lorentz invariant bulk-boundary propagator in detail for the $\tmop{BMS}_3$/$\tmop{CCFT}_2$ case. In our calculation, it is necessary to introduce a nonzero mass at the beginning and take it as vanishing at the end. The framework we proposed has the potential to probe local bulk physics from the boundary CCFT. 
\end{abstract}

\vfill{\footnotesize E-mail: bchen01@pku.edu.cn, z.z.hu@pku.edu.cn}

\newpage

\section{Introduction}

One essential problem in AdS/CFT is to understand the emergence of bulk physics from boundary CFT. (For a nice review on this subject, see \cite{Harlow:2018fse}.) It is important to understand how the bulk local operators can be represented by the boundary operators. This so-called bulk reconstruction in $\tmop{AdS}/\tmop{CFT}$ has been explored for many years. For the early efforts, see \cite{Banks:1998dd,Balasubramanian:1998sn,Bena:1999jv}. In \cite{PhysRevD.74.066009},  A. Hamilton et.al. proposed to use a smearing function as a bulk-boundary kernel to express a bulk field as an integral of the dual boundary field.  In fact, the integral of the dual boundary field can be expanded and simply becomes a linear combination of a series of descendants. The HKLL proposal was further investigated to understand the bulk dynamics\cite{Kabat:2011rz, Heemskerk:2012mn,Kabat:2013wga,Morrison:2014jha,Kabat:2015swa}. A few years ago, another reconstruction of local bulk states and their corresponding operators was proposed by using the Ishibashi state in dual CFT\cite{Miyaji:2015fia,Ishibashi:1988kg,Verlinde:2015qfa,Nakayama:2015mva}. When considering the graviton propagating between the fields in $\tmop{AdS}_3$, N. Anand et.al. \cite{Anand:2017dav} further expressed the scalar bulk field as a linear combination of Virasoro descendants and checked the consistency in the large $c$ limit \cite{Chen:2017dnl}.

In the past decade,  there have been lots of studies on holography in flat spacetime. As the asymptotic symmetry group of asymptotically flat spacetime is the Bondi-Metzner-Sachs (BMS) group\cite{Bondi:1962px,Sachs:1962wk}, it has been widely believed that gravity in asymptotically flat spacetime could be holographically dual to a field theory with BMS symmetry\cite{Barnich:2010eb,Bagchi:2010zz, Bagchi:2012cy,Bagchi:2012xr,Barnich:2012xq}. The BMS symmetry is non-relativistic, isomorphic to Carrollian conformal symmetry\cite{Duval:2014uva}. For other studies on flat holography in three dimensions, see \cite{Barnich:2012rz,Barnich:2013yka,Barnich:2014kra,Bagchi:2014iea,Campoleoni:2016vsh,Bagchi:2016geg,Bagchi:2017cpu,Jiang:2017ecm,Hijano:2017eii,Fuentealba:2017omf,Hijano:2018nhq,Bagchi:2019unf,Hijano:2019qmi,Merbis:2019wgk,Chen:2020vvn, Ammon:2020wem,Chen:2022cpx,Chen:2022jhx}.

It turns out that the Carrollian conformal symmetry is closely related to celestial holography\cite{Chen:2021xkw,Donnay:2022aba, Bagchi:2022emh, Donnay:2022wvx,Bagchi:2023fbj,Donnay:2023mrd,Saha:2023hsl,Saha:2023abr,Mason:2023mti}, another realization of flat holography. The celestial holography and $\tmop{BMS}_4$/$\tmop{CCFT}_3$ approaches are related by (Modified) Mellin transformations \cite{Donnay:2022aba,Bagchi:2022emh,Donnay:2022wvx}. However there are a few subtleties and puzzles in flat holography. For example, as massive particles can never reach the null infinity, it seems that one should not consider the massive fields in the bulk and correspondingly one should only focus on the $\xi=0$ sector in dual field theory. Another perplexing issue is the representations of Carrollian conformal algebra. Considering the flat space as the large radius limit of AdS spacetime would lead to the conclusion that the field in flat space is related to the operator in the induced representation in the dual boundary CCFT \cite{Barnich:2014kra,Hao:2021urq} from which it is hard to establish the operator-state correspondence.


The bulk reconstruction in flat holography has been studied to some extent in the literature. In \cite{Hijano:2019qmi}, the HKLL proposal in flat spacetime has been found by taking the flat space limit of AdS spacetime, and the induced representation in the dual field theory has been used to compute the correlators\footnote{For another route to the bulk reconstruction of conformally coupled fields in flat space, see \cite{Bhattacharyya:2023czi}.}. In the very recent works including \cite{Donnay:2022wvx,Nguyen:2023vfz,Nguyen:2023miw}, it has been proposed that the free field can be reconstructed from the highest-weight representation in the CCFT as well. In particular, the bulk free field has been reconstructed via a bulk-boundary propagator by using the Kirchhoff-d’Adhemar formula\cite{Penrose:1980yx} in the context of $\tmop{BMS}_4$/$\tmop{CCFT}_3$\cite{Donnay:2022wvx}.

In this note, we revisit the bulk reconstruction in flat holography. We take an algebraic way by using boundary operator expansion to represent a bulk operator. However, due to the weird behavior of the two-point functions in the boundary CCFT for the massless case, we will start with considering the bulk reconstruction for a massive field from the highest-weight representation of CCFT. As we will show, even though the Poincar\'e symmetry of the massive bulk field we reconstructed is broken, it gets restored after taking the massless limit. Finally, we manage to obtain the bulk-boundary propagator from the boundary operator expansion, which is exactly the same as the one by directly using the explicit Kirchhoff-d’Adhemar formula shown in \cite{Donnay:2022wvx}.

The remaining parts of the note are organized as follows. In section \ref{section2}, we study the scalar and also vector bulk reconstructions in $\tmop{BMS}_3$/$\tmop{CCFT}_2$. In section \ref{section3}, we investigate the scalar bulk reconstruction in  $\tmop{BMS}_4$/$\tmop{CCFT}_3$. We conclude in the section \ref{section4}.


\section{Bulk reconstruction in $\tmop{BMS}_3$/$\tmop{CCFT}_2$}\label{section2}

\subsection{$\tmop{BMS}_3$ symmetry}\label{section2-1}

In Minkowski spacetime $\mathbb{R}^{1, 2}$ , we can introduce a new coordinate system $(r,u,\phi)$ as
\be
  X^{\mu}  =   (0, r\phi, 0) +(u+\frac{r\phi^2}{2}) (1,0,-1)+\frac{r}{2}(1,0,1)\,, \hs{3ex}\m=0,1,2,
\ee
in terms of which the metric becomes
\be
  d s^2  =  \eta_{\mu \nu} d X^{\mu} d X^{\nu} = - 2 d u d r + r^2 d \phi^2\,.
\ee
We can obtain the null infinity, which is a null plane where a $\tmop{CCFT}_2$ can live, by taking $r\rightarrow\infty$. Any interval on this null plane can be null or spacelike, and this is why a massive particle with a timelike worldline can never reach the null infinity.

The Carrollian conformal transformations on the null boundary are generated by the $\tmop{BMS}_3$ algebra, whose generators are the vectors
\be
M_{n} = -\phi^{n+1}
\partial_u\,, \quad L_m = - \phi^{m + 1} \partial_\phi -(m+1) \phi^{m}u \partial_u\,,\hs{3ex}m,n\in \mathbb{Z},
\ee
satisfying the following commutation relations
\be
\bag
  &[L_m, L_n]  =  (m - n) L_{m + n}\,,\\
  &[L_m, M_n]  =  (m-n) M_{m+n}\,,\\
  &[M_m, M_n] =0.
\eag
\ee
The global part of Carrollian conformal symmetry is simply the Poincar\'e group\footnote{All the results of this work involve only Poincar\'e algebra, the global part of the BMS$_3$ algebra, but some discussions on the full BMS$_3$ may be useful for future study.}, generated by $M_{0}$, $M_{\pm}$, $L_{0}$, $L_{\pm}$,
\be\label{casirela2}
\bag
 &M_{-1} = -\partial_0+\partial_2\,,\\
 &M_{0}  =  \partial_1\,,\\
&M_{1} = -(\partial_0+\partial_2)\,,\\
\eag
\bag
 & \qquad L_{- 1}  =  J_{01}+J_{12}\,,\\
 & \qquad L_0 =  -J_{02}\,,\\
 & \qquad L_1 =  -J_{01}+J_{12}\,,
\eag
\ee
where we have defined that $\partial_\mu \equiv \partial_{X^\mu}$ and $J_{\mu \nu} \equiv X_{\mu} \partial_{\nu} - X_{\nu} \partial_{\mu}\,.$
Using the coordinate transformation relations,
we have
\be\label{tran2}
\bag
  &M_{-1}  =   -\partial_u\,,\qquad\\
  &M_{ 0}  =  -\phi \partial_u + \frac{1}{r} \partial_\phi\,,\qquad\\
  &M_{1}  =  -\phi^2\partial_u-2\partial_r+2\frac{\phi}{r}\partial_\phi\,,\qquad\\
\eag
\bag
  &L_{- 1}  =  - \partial_\phi\,,\\
  &L_0  =  -\phi\partial_\phi-u\partial_u+r\partial_r\,,\\
  &L_1  = - \phi^2 \partial_\phi + 2\phi (r \partial_r - u \partial_u) + \frac{2u}{r}\partial_\phi\,.
\eag
\ee
As we want a bulk field in correspondence with a highest-weight representation of CCFT$_3$, we need to define the Hermitian conjugation of these generators to be
\be\label{conju}
  L_m^{\dagger}  =  L_{- m},\hs{3ex}
  M^{\dagger}_m  =  M_{-m}\,.
\ee
Moreover, we need the inversion transformation \cite{Hao:2021urq} associated to the Hermitian conjugation relation (\ref{conju}) in the boundary $\tmop{CCFT}_2$,
\be
\bag
  \phi  & \rightarrow \phi^\prime =\frac{1}{\phi}\,,\\
  u  & \rightarrow u^\prime =-\frac{u}{\phi^2}\,,\\
\eag
\ee
and thereafter the transformation rule for a primary operator becomes
\be
\mathcal{O}^\dagger(u,\phi)=\phi^{-2\Delta} \exp{-2\xi\frac{u}{\phi}}\mathcal{O}(-\frac{u}{\phi^2},\frac{1}{\phi})\,.
\ee

\subsection{Scalar bulk reconstruction in $\tmop{BMS}_3$/$\tmop{CCFT}_2$}\label{section2-2}

To reconstruct a free scalar in the bulk, we start by considering what kind of conditions the primary operator $\mathcal{O}$ should satisfy. Firstly a primary state in CCFT$_2$ is characterized by scaling dimension $\D$ and boost charge $\xi$\cite{Bagchi:2009pe,Bagchi:2009ca,Chen:2020vvn},
\be
[L_0, \mathcal{O}(0)] =\Delta \mathcal{O}(0), \hs{3ex} [M_{0},  \mathcal{O}(0)]   =  \xi  \mathcal{O}(0).
\ee
 From the $\tmop{BMS}_3$ algebra, we have
\be
\bag
[L_0,L_m]&=-m L_m\,,\\
[L_0,M_n]&=-n M_n\,.
\eag
\ee
Corresponding to the highest-weight representation, the generators lowering the scaling dimension should annihilate the primary state,
\be
\bag
  &[L_m, \mathcal{O}(0)] =  0 \quad\text{when}\quad m >0\,,\\
  &[M_n, \mathcal{O}(0)]  =  0 \quad\text{when}\quad n > 0\,.
\eag
\ee
Secondly, the mass of the field is given by the Casimir operators $\partial^2$, which commute with all the generators of Poincar\'e group,
\be\label{casi}
  [\partial^2, \mathcal{O}(0)]  =  m^2 \mathcal{O}(0)\,,
\ee
where $m^2$ is the square of the mass of the scalar bulk field. Since
\be\label{casim}
  \partial^2  =  -  M_{1} M_{-1} +  M_{0}^2\,,
\ee
we have
\be
m^2=\xi^2.
\ee

Now we can construct the bulk field $\phi(r,u,\phi)$ as
\be
\phi (r, 0, 0) | 0 \rangle \nobracket \nobracket = \sum^\infty_{J,
L=0} \lambda_{J, L} r^{- K} L_{- 1}^J  M_{-1}^L | \mathcal{O}\rangle \nobracket
\nobracket\,.
\ee
Since the bulk field is a scalar, it must transform under Poincar\'e group in the following way, $[Q, \phi (X^{\mu})] = - \delta_Q \phi (X^{\mu})$, where $Q$ can be any
 conserved charge corresponding to the transformation in Poincar\'e group and $\delta_Q$ is given by (\Ref{tran2}). More explicitly, we have
\be\label{tran3}
\bag
  L_0 \phi (r, 0, 0) | 0 \rangle \nobracket \nobracket & =  -  r
  \partial_r \phi (r, 0, 0) | 0 \rangle \nobracket \nobracket \,,\\
  L_1 \phi (r, 0,0) | 0 \rangle \nobracket \nobracket & =  0 \,,\\
  M_1 \phi (r, 0, 0) | 0 \rangle \nobracket \nobracket & =  2\partial_r
  \phi (r, 0, 0) | 0 \rangle \nobracket \nobracket \,,\\
  \left( M_{0} + \frac{1}{r} L_{- 1} \right) \phi (r, 0, 0) | 0 \rangle
  \nobracket \nobracket & =  0 \,.
\eag
\ee
The first equation in (\Ref{tran3}) tells
\be
  K =  \Delta + J + L \,.
\ee
The second equation in (\Ref{tran3}) tells
\be
  2\xi (L + 1) \lambda_{J, L + 1} + (J+2L +2\Delta) (J +
  1) \lambda_{J + 1, L}  =  0 \,.\\
\ee
The third equation in (\Ref{tran3}) tells
\be\label{eq44}
  2(J+1)\xi\lambda_{J+1, L} + (J+2)(J+1) \lambda_{J+2 , L-1} =  -2(\Delta+J+L)\lambda_{J, L} \,.
\ee
The fourth equation in (\Ref{tran3}) tells
\be\label{eq44}
  (J+1)\lambda_{J+1, L-1} + \xi \lambda_{J , L} +
  \lambda_{J - 1, L }  =  0 \,.
\ee
Solving the first three equations, we can obtain a general iteration relation for $\lambda_{J,L}$ as
\be
\bag
  \lambda_{J + 1, L} & =  -\frac{1}{\xi}  \frac{(J + 2L + 2\Delta-1)(J+L+\Delta)}{(J + 1)(J+L+2\Delta-1)}
  \lambda_{J, L}\,,\\
  \lambda_{J, L + 1} & = \frac{1}{\xi^2}  \frac{(J + 2L + 2\Delta-1)(J+2L + 2\Delta)(J+L+\Delta)}{2(L + 1)(J+L+2\Delta-1)}
  \lambda_{J, L}\,.\\
\eag
\ee
Setting $\lambda_{0,0}=1$, it can be expressed as
\be\label{recur2}
\lambda_{J, L}=\frac{2^{-L}\left(\frac{1}{\xi^2}\right)^L \left(-\frac{1}{\xi}\right)^J \Gamma(J+L+\Delta) \Gamma(J+2L+2\Delta-1)}{\Gamma(J+1)\Gamma(L+1)\Gamma(\Delta)\Gamma(J+L+2\Delta-1)}\,.
\ee
It is clear that the solution does not make sense if $\xi$ is vanishing.
However, if we keep $\xi$ non-vanishing, the solution does not satisfy the fourth equation. In fact, using this solution, the fourth equation in (\Ref{tran3}) becomes
\be
\left( M_{0} + \frac{1}{r} L_{- 1} \right) \phi (r, 0, 0) | 0 \rangle
  \nobracket \nobracket  =  \frac{1}{r} (\Delta-1)\xi\int_{r}^{\infty}d r \phi(r,0,0)| 0 \rangle\quad\text{for} \quad\Delta\neq 1\,,
\ee
and
\be\label{simpcase32}
\left( M_{0} + \frac{1}{r} L_{- 1} \right) \phi (r, 0, 0) | 0 \rangle
  \nobracket \nobracket  =  \frac{\lambda_{0,0}\xi}{r^\Delta}\mathcal{O}(0,0)| 0 \rangle\quad\text{for} \quad\Delta= 1\,.
\ee
In any case, the equation could be reduced to the fourth equation in (\Ref{tran3}) only if $\xi=0$. It seems that the Poincar\'e symmetry is broken except for the massless case. The role of a non-vanishing mass (or $\xi$) might be regarded as a regulator, and our regulation scheme somewhat breaks the Poincar\'e global symmetry. We will follow the prescription that we do computation with a non-vanishing $\xi$ and take $\xi=0$ at the end to restore the Poincar\'e symmetry.

So far we have reconstructed a free scalar from a highest-weight representation of $\tmop{CCFT}_2$. And the bulk field at a general point $(r, u, z,\bar{z})$ is obtained by
\be
  \phi (r, u, \phi) = \exp (u M_{-1}+\phi L_{-1}) \phi (r, 0, 0) \exp (-u M_{-1}-\phi L_{-1})\,.
\ee
Its action on the vacuum is
\be\label{expan}
  \phi (r, u, \phi) | 0 \rangle \nobracket \nobracket 
  =  \sum_{J, L} \lambda_{J, L} r^{- (\Delta + J + L)} e^{\phi L_{-1}+u M_{-1}} L_{- 1}^J
  M_{-1}^L | \mathcal{O}\rangle \nobracket \nobracket\,.
\ee

The above reconstruction of bulk scalar field can be rewritten in terms of the boundary operator, 
\be\label{boe}
\phi(r,0,0)=\sum^\infty_{J,L=0} \lambda_{J, L} r^{- K} \partial_{\phi}^J  \partial_{u}^L \mathcal{O}(0,0)\,. 
\ee

\subsection{Bulk-boundary scalar propagator in $\tmop{BMS}_3$/$\tmop{CCFT}_2$}\label{section2-3}

We denote the bulk-boundary propagator for a free field as
\be
G^\xi(r,u,\phi)\equiv\langle\phi(r,u,\phi)\mathcal{O}(0,0)\rangle\,,
\ee
where $\xi$ indicates the mass of the field. When considering the massless case, the bulk-boundary propagator is determined by the differential equation
\be
\partial^2 G^{\xi=0}(r,u,\phi)=-\frac{1}{r}\partial_r\left(r\partial_r G\right)-\frac{1}{r}\partial_u\left(r\partial_r G\right)+\frac{1}{r}\partial_\phi\left(\frac{1}{r}\partial_\phi G\right)=0\,,\ee
with the boundary condition
\be
G^{\xi=0}(r,u,\phi)\sim\frac{1}{r^\Delta} \langle \mathcal{O}(u,\phi)\mathcal{O}(0,0)\rangle=\frac{1}{r^\Delta}\frac{1}{\phi^{2\Delta}}\,.
\ee
The solution is 
\be
G^{\xi=0}(r,u,\phi)=\left(\frac{1}{r\phi^2+2u}\right)^\Delta\,. \label{bulkBcorrelator}
\ee
This method can be further used to obtain the bulk-bulk propagators.
Considering the Poincar\'e symmetry, the bulk-bulk propagator must be of the form
\be
G^{\xi=0}(X_1,X_2)=\left(\frac{1}{\eta_{\mu\nu}(X_1-X_2)^\mu(X_1-X_2)^\nu}\right)^\Delta\,.
\ee
Then the equation $\partial^2 G^{\xi=0}(X_1,X_2)=0$ fixes the scaling dimension $\Delta=\frac{1}{2}$. 

We would like to reproduce the bulk-boundary propagator \eqref{bulkBcorrelator} by using the reconstructed bulk field \eqref{boe}. Before we do explicit computation, we need to clarify from a different point of view why we cannot set the mass (or $\xi$) to zero at the very first beginning of the computation. Assuming the expression for $\lambda_{J,L}$ has been obtained, we substitute (\ref{boe}) into the bulk-boundary propagator as
\be
\bag
\langle\mathcal{O}(u,\phi)\phi(r,0,0) \rangle=&\sum^\infty_{J,L=0} \lambda_{J, L} r^{- K} (-\partial_{\phi})^J  (-\partial_{u})^L \langle\mathcal{O}(u,\phi)O(0,0) \rangle\\
=&\sum^\infty_{J,L=0} \lambda_{J, L} r^{- K} (-\partial_{\phi})^J  (-\partial_{u})^L \left(\frac{A}{\phi^{2\Delta}}+\frac{B \delta(\phi)}{u^{2\Delta-1}}\right)\,,\\
\eag
\ee
then there is no way that we can get any nontrivial result: as the first term on the right-hand side does not depend on $u$ while the second term is singular for $\phi$, the final result is singular, not a well-behaved function of $u$ and $\phi$. On the other hand for a nonzero mass (or $\xi$), it becomes
\be\label{summ}
\langle\mathcal{O}(u,\phi)\phi(r,0,0) \rangle=\sum^\infty_{J,L=0} \lambda_{J, L} r^{- K} (-\partial_{\phi})^J  (-\partial_{u})^L \left(A\phi^{-2\Delta}\exp{-2\xi\frac{u}{\phi}}\right)\,,\\
\ee
which may lead to a well-behaved function. In the above discussion, we have used the fact that the two-point functions of primary operators in CCFT take the form\cite{Bagchi:2009ca,Bagchi:2009pe,Chen:2022jhx} 
\be
\langle\mathcal{O}(u,\phi)\mathcal{O}(0,0) \rangle=\left\{\begin{array}{cc}
 A\phi^{-2\Delta}+B \delta(\phi)u^{-(2\Delta-1)},    & \mbox{for $\xi=0$,} \\
 A\phi^{-2\Delta}\exp{-2\xi\frac{u}{\phi}},    & \mbox{for $\xi\neq 0$.}
\end{array}\right.
\ee

From now on, we will derive the bulk-boundary propagator result in $\tmop{BMS}_3$/$\tmop{CCFT}_2$. The most direct but kind of clumsy way to calculate is just using the equation (\ref{summ}),
\be
\bag
\langle\mathcal{O}(u,\phi)\phi(r,0,0) \rangle=&\sum^\infty_{J,L=0} \lambda_{J, L} r^{- (J+L+\Delta)} (-\partial_{\phi})^J  (-\partial_{u})^L \left(\phi^{-2\Delta}\exp{-2\xi\frac{u}{\phi}}\right)\\
=&\sum_{M=0}^\infty \sum_{J=0}^M \lambda_{J, M-J} r^{- (M+\Delta)} (-\partial_{\phi})^J  (-\partial_{u})^{M-J} \left(\phi^{-2\Delta}\exp{-2\xi\frac{u}{\phi}}\right)\\
=&\exp{-2\xi \frac{u}{\phi}}\left(\frac{1}{r\phi^2-2u}\right)^\Delta\,.
\eag
\ee
Another much more sophisticated way is to consider
\be\label{soph}
\bag
\langle0|\mathcal{O}(u,\phi\text{)}\phi\text{(}r,0,0\text{)}|0\rangle
=& \left\{\phi^{-2\Delta}e^{-2\xi\frac{u}{\phi}}\mathcal{O}\left(-\frac{u}{\phi^{2}},\frac{1}{\phi}\right)|0\rangle\right\}^{\dagger}\phi(r,0,0)\mid0\rangle   \\
=&\left\{\phi^{-2\Delta}e^{-2\xi\frac{u}{\phi}}e^{-\frac{u}{\phi^{2}}M_{-1}+\frac{1}{\phi}L_{-1}}\mathcal{O}\left(0,0\right)\mid0\rangle\right\}^{\dagger}\phi(r,0,0)\mid0) \\
=&\phi^{-2\Delta}e^{-2\xi\frac u\phi}\langle \mathcal{O}|e^{-\frac u{\phi^2}M_1+\frac1\phi L_1}\phi(r,0,0)|0\rangle\,,  \\
\eag
\ee
then using the second and the third equations in (\ref{tran3}), it becomes
\be\label{secequality}
\bag
\phi^{-2\Delta}e^{-2\xi\frac u\phi}\langle \mathcal{O}|e^{-\frac u{\phi^2}M_1+\frac1\phi L_1}\phi(r,0,0)|0\rangle
=&\phi^{-2\Delta}e^{-2\xi\frac{u}{\phi}}e^{-\frac{u}{\phi^{2}}2\partial_{r}}\langle \mathcal{O}|\phi(r,0,0)|0\rangle \\
=&\phi^{-2\Delta}e^{-2\xi\frac u\phi}e^{-\frac u{\phi^2}2\partial_r}r^{-\Delta}  \\
=&\exp{-2\xi \frac{u}{\phi}}\left(\frac{1}{r\phi^2-2u}\right)^\Delta\,.
\eag
\ee
Finally, we have derived the  bulk-boundary propagator in the presence of the non-vanishing mass, 
\be
G^\xi(r,u,\phi)=\exp{-2\xi \frac{u}{\phi}}\left(\frac{1}{r\phi^2+2u}\right)^\Delta\,.
\ee
And it reduces to the massless case when simply setting the mass (or $\xi$) to zero.

\subsection{Vector bulk reconstruction in $\tmop{BMS}_3$/$\tmop{CCFT}_2$}\label{section2-4}
We consider a massive vector field in the bulk, which can be described by the action
\be
\mathcal{L}=-\frac{1}{4}F^{\mu\nu}F_{\mu\nu}-\frac{1}{2}m^2 A^\mu A_\mu\,.
\ee
This model can be realized as a U(1) symmetry-broken gauge theory. The equations of motion for the vector field $A_\mu$ are
\be
\partial_\mu A^\mu=0\,,\qquad (-\partial^2+m^2)A_{\mu}=0\,.
\ee
The vector field has two physical degrees of freedom. Without losing generality, we choose them to be  $\{A^u, A^\phi\}$, in terms of which the component $A^r$ can be expressed as
\be
A^r=\int^{\infty}_{r} (\partial_u A^u+\partial_\phi A^\phi)d r\,.
\ee

There is more than one physical degree of freedom in the bulk field theory, thus there may exist more than one conformal family appearing in the reconstruction. For a specific conformal family, we make the ansatz
\be
\bag
A^u (r, 0, 0) | 0 \rangle \nobracket \nobracket &= \sum^\infty_{J,
L, M=0} U_{J, L} r^{- K} L_{- 1}^J M_{-1}^L | \mathcal{O}^u\rangle \nobracket
\nobracket\,,\\
A^\phi (r, 0, 0) | 0 \rangle \nobracket \nobracket &= \sum^\infty_{J,
L, M=0} \Phi_{J, L} r^{- K} L_{- 1}^J M_{-1}^L | \mathcal{O}^\phi\rangle \nobracket
\nobracket\,.
\eag
\ee
Then the Poincar\'e symmetry transformations on the field, $[Q, A^i (X^{\mu})] = - \delta_Q A^i (X^{\mu})$, are almost the same except that we should replace the ordinary derivatives with Lie-derivatives,
\be\label{A5}
\bag
[L_0, A^u(r,0,0)]&=-r\partial_r A^u(r,0,0)-A^u(r,0,0)\,,\\
[L_0, A^\phi(r,0,0)]&=-r\partial_r A^\phi(r,0,0)-A^\phi(r,0,0)\,,\\
[L_0, A^r(r,0,0)]&=-r\partial_r A^r(r,0,0)+A^r(r,0,0)\,,\\
\eag
\ee
\be\label{A6}
\bag
[L_1, A^u(r,0,0)]&=0\,,\\
[L_1, A^\phi(r,0,0)]&=\frac{2}{r}A^u\,,\\
[L_1, A^r(r,0,0)]&=2r A^{\phi}(r,0,0)\,,\\
\eag
\ee
\be\label{A7}
\bag
[M_1, A^u(r,0,0)]&=2\partial_r A^u(r,0,0)\,,\\
[M_1, A^\phi(r,0,0)]&=2\partial_r A^\phi(r,0,0)+\frac{2}{r}A^\phi(r,0,0)\,,\\
[M_1, A^r(r,0,0)]&=2\partial_r A^r(r,0,0)\,,\\
\eag
\ee
\be
\bag
[(M_{0}+\frac{1}{r}L_{-1}), A^u(r,0,0)]&=-A^{\phi}(r,0,0)\,,\\
[(M_{0}+\frac{1}{r}L_{-1}), A^\phi(r,0,0)]&=-\frac{1}{r^2}A^r(r,0,0)\,,\\
[(M_{0}+\frac{1}{r}L_{-1}), A^r(r,0,0)]&=0\,.\\
\eag
\ee
Substituting the ansatz into (\ref{A5}), we obtain
\be
K=J+L+\Delta+1\,.
\ee
The difference of the expression for $K$ is induced by the spin of the massless vector field.

The full computation of the bulk two-point functions could be quite complicated. But we can  obtain bulk-boundary propagator directly, like in \eqref{soph}, and obtain
\be\label{A10}
\langle A^u(r,u,\phi)\mathcal{O}^u(0,0)\rangle=\exp{-2\xi \frac{u}{\phi}}\frac{\phi^2}{\left(r\phi^2+2u\right)^{\Delta+1}}\,.
\ee
Here we have used the behavior of $A^u$ in (\ref{A6}) and (\ref{A7})  in the derivation of this propagator. After taking the limit $\xi\to 0$, we see that 
\be
\langle A^u(r,u,\phi)\mathcal{O}^u(0,0)\rangle=\frac{\phi^2}{\left(r\phi^2+2u\right)^{\Delta+1}}\,.
\ee
We can also obtain all the bulk-boundary propagators by directly considering how the Poincare symmetry constrains the massless bulk-bulk propagator, and find
\be
\langle A^{\mu}(X_1)A^{\nu}(X_2)\rangle=\frac{(X_1-X_2)^{\mu}(X_1-X_2)^{\nu}+c_n \eta^{\mu\nu}\Lambda}{\Lambda^{n+1}}
\ee
with
\be
\Lambda=\eta_{\mu\nu} (X_1-X_2)^\mu(X_1-X_2)^\nu\,,
\ee
where the coefficient $c_n=(1-n)/n$ is required to maintain the condition $\partial_\mu \langle A^{\mu} A^{\nu}\rangle=0$.
And the equation $\partial^2 \langle A^{\mu} A^{\nu}\rangle=0$ would give $n=3/2$. Then we  can check that the propagator
\be
\bag
\langle A^u(r,u,\phi) A^u(r',0,0)\rangle &=\frac{4n u^2+2(2-n)(r-r')u\phi^2-(2-n)r r'\phi^4}{4n\left(-2u(r-r')+r r'\phi^2\right)^{n+1}}\\
&\sim -\frac{1}{(r')^n} \frac{(2-n)\phi^2}{4n(r \phi^2+2u)^n}
\eag
\ee
has the similar form with (\ref{A10}) when considering its asymptotic large $r'$ behavior.

\section{Bulk reconstruction in $\tmop{BMS}_4/\tmop{CCFT}_3$}\label{section3}

\subsection{$\tmop{BMS}_4$ symmetry}\label{section3-1}
In Minkowski spacetime $\mathbb{R}^{1, 3}$, we can introduce a new coordinate system $(r,u,z,\bar{z})$ as
\be
  X^{\mu}  =  \frac{1}{\sqrt{2}} u (1, 0, 0, - 1) + \frac{1}{\sqrt{2}} r (1
  + z \bar{z}, z + \bar{z}, - i (z - \bar{z}), 1 - z \bar{z}),\qquad\mu=0,1,2,3 \,,
\ee
in terms of which the metric can be rewritten as 
\be
  d s^2  =  \eta_{\mu \nu} d X^{\mu} d X^{\nu} = - 2 d u d r + 2 r^2 d z d
  \bar{z}\,.
\ee
We can obtain the null infinity, where a $\tmop{CCFT}_3$ lives, by taking the limit $r\rightarrow\infty$.

The Carrollian conformal transformation on the null infinity is generated by the  vectors
\be
\bag
&M_{r, s} = - z^r \bar{z}^s
\partial_u\,, \\
 &L_m = - z^{m + 1} \partial -\frac{1}{2}(m+1) z^{m}u \partial_u\,,\\
&\bar{L}_m = - \bar{z}^{m + 1} \bar{\partial} -\frac{1}{2}(m+1) \bar{z}^{m}u \partial_u\,,
\eag
\ee
where $\partial$ stands for $\partial_z$ and $\bar{\partial}$ stands for $\partial_{\bar{z}}$, obeying the $\tmop{BMS}_4$ algebra\footnote{The BMS$_4$ algebra takes a different form from the one in \cite{Chen:2021xkw}, but they are related to each other.}
\be
\bag
  &[L_m, L_n]  =  (m - n) L_{m + n}\,,\\
  &[\bar{L}_m, \bar{L}_n]  =  (m - n) \bar{L}_{m + n}\,,\\
  &[L_m, M_{r, s}]  =  \left( \frac{m + 1}{2} - r \right) M_{r + m, s}\,,\\
  &[\bar{L}_m, M_{r, s}] =  \left( \frac{m + 1}{2} - s \right) M_{r , s+m}\,.
\eag
\ee
The global symmetry is the Poincar\'e group, generated by $M_{0, 0}$, $M_{1, 0}$, $M_{0,
1}$, $M_{1, 1}$, $L_{\pm, 0}$ and $\bar{L}_{\pm, 0}$. Their actions on the bulk point are given by
\be\label{casirela}
\bag
 &M_{0, 0}  =  \frac{- \partial_0 + \partial_3}{\sqrt{2}}\,,\\ &M_{1, 1} =  \frac{- \partial_0 - \partial_3}{\sqrt{2}}\,,\\
 &M_{1, 0}  =  \frac{\partial_1 + i \partial_2}{\sqrt{2}}\,,\\ &M_{0, 1} =  \frac{\partial_1 - i \partial_2}{\sqrt{2}}\,,\\
\eag
\bag
 & \qquad L_{- 1}  =  \frac{1}{2} (- J_{10} + i J_{20} - i J_{23} + J_{13})\,,\\
 & \qquad \bar{L}_{- 1}  =  \frac{1}{2} (- J_{10} - i J_{20} + i J_{23} + J_{13})\,,\\
 & \qquad L_1 =  \frac{1}{2} (J_{10} + i J_{20} + i J_{23} + J_{13})\,,\\
 & \qquad \bar{L}_1 =  \frac{1}{2} (J_{10} - i J_{20} - i J_{23} + J_{13})\,,\\
 & \qquad L_0 =  \frac{1}{2} (- J_{03} + i J_{12})\,,\\
 & \qquad \bar{L}_0 =  \frac{1}{2} (- J_{03} - i J_{12})\,,\\
\eag
\ee
where we have defined that $\partial_\mu \equiv \partial_{X^\mu}$ and $J_{\mu \nu} \equiv X_{\mu} \partial_{\nu} - X_{\nu} \partial_{\mu}\,.$
Using the coordinate transformation relations, 
we have
\be\label{tran}
\bag
  &M_{0, 0}  =  - \partial_u\,,\qquad\\
  &M_{1, 0}  =  - z \partial_u + \frac{1}{r} \bar{\partial}\,,\qquad\\
  &M_{0, 1}  =  - \bar{z} \partial_u + \frac{1}{r} \partial\,,\qquad\\
  &M_{1, 1}  =  - z \bar{z} \partial_u - \partial_r + \frac{1}{r} (z \partial
  + \bar{z} \bar{\partial})\,,\qquad\\
\eag
\bag
  &L_{- 1}  =  - \partial\,,\\
  &\bar{L}_{- 1}  =  - \bar{\partial}\,,\\
  &L_0  =  \frac{1}{2} (r \partial_r - u \partial_u - 2 z \partial)\,,\\
  &\bar{L}_0  =  \frac{1}{2} (r \partial_r - u \partial_u - 2 \bar{z} \bar{\partial})\,,\\
  &L_1  = - z^2 \partial + z (r \partial_r - u \partial_u) + \frac{u}{r}
  \bar{\partial}\,,\\
  &\bar{L}_1  = - \bar{z}^2 \bar{\partial} + \bar{z} (r \partial_r - u \partial_u) + \frac{u}{r}
  \partial\,.
\eag
\ee

\subsection{Scalar bulk reconstruction in $\tmop{BMS}_4$/$\tmop{CCFT}_3$}\label{section3-2}

To reconstruct a massive free scalar in the bulk, we start with the primary operator $\mathcal{O}$ in boundary CCFT. Firstly, from the BMS$_4$ algebra, we have
\be
\bag
[L_0+\bar{L}_0,L_m]&=-m L_m\,,\\
[L_0+\bar{L}_0,\bar{L}_m]&=-m \bar{L}_m\,,\\
[L_0+\bar{L}_0,M_{r,s}]&=(1-r-s)M_{r,s}\,.
\eag
\ee
A highest-weight representation is defined by its scaling dimension $\Delta$ and "spin" $S$, 
\be
\bag
L_0| \mathcal{O}\rangle \nobracket \nobracket&=h | \mathcal{O}\rangle \nobracket \nobracket\,,\qquad h=\frac{\Delta+S}{2}\,,\\
\bar{L}_0| \mathcal{O}\rangle \nobracket \nobracket&=\bar{h} | \mathcal{O}\rangle \nobracket \nobracket\,,\qquad \bar{h}=\frac{\Delta-S}{2}\,.
\eag
\ee
and the annihilation conditions
\be
\bag
  [L_m, \mathcal{O}(0)] & =  0, \quad\text{when}\quad m >0\,,\\
  [\bar{L}_m, \mathcal{O}(0)] & =  0, \quad\text{when}\quad m >0\,,\\
  [M_{r, s}, \mathcal{O}(0)] & =  0, \quad\text{when}\quad r + s > 1\,.
\eag
\ee 
The "spin" $S$ is, in fact, irrelevant in the bulk reconstruction, because we can always define a new primary state (thereafter a new highest-weight representation) with no "spin" as
\be
| \tilde{O} \rangle \nobracket \nobracket=M_{1,0}^S| \mathcal{O}\rangle \nobracket \nobracket\,,
\ee
so in the following discussions, we simply choose the highest weight representation with "spin" $S$ equals zero. Actually, the spin of the operator in CCFT is a little bit tricky, and it should be defined in a more sophisticated way\cite{Chen:2021xkw}.


Secondly, the mass of the field is given by the Casimir operator $\partial^2$, which can be translated into $(r,u,z,\bar{z})$ coordinate system by using the relations (\ref{casirela}). Since the Casimir operators commute with any Poincar\'e group generators, which give the descendants by acting on the primary states,  we must have
\be\label{casi}
  [\partial^2, \mathcal{O}(0)]  =  m^2 \mathcal{O}(0)\,,
\ee
with $m^2$ being the square of the mass of the scalar bulk field. In BMS$_4$, as the Casimir operator 
\be\label{casim}
  \partial^2  =  - 2 M_{0, 0} M_{1, 1} + 2 M_{0, 1} M_{1, 0}\,,
\ee
we have
\be\label{mass22}
  2 M_{0, 1} M_{1, 0} | \mathcal{O}\rangle \nobracket \nobracket  =  m^2 | \mathcal{O}\rangle
  \nobracket \nobracket\,,
\ee
from which we can define that
\be
   \left(\frac{\sqrt{2} M_{1, 0}}{m} \right)^{- 1} | \mathcal{O}\rangle \nobracket \nobracket=\frac{\sqrt{2} M_{0, 1}}{m} | \mathcal{O}\rangle \nobracket \nobracket \,.
\ee

For a scalar bulk field, it must transform under Poincar\'e group in an appropriate way. More precisely, it must obey the Ward identity $[Q, \phi (X^{\mu})] = - \delta_Q \phi (X^{\mu})$, where $Q$ can be any
Poincar\'e generator and $\delta_Q$ is given by (\Ref{tran}), that is
\be\label{tran1}
\bag
  L_0 \phi (r, 0, 0,0) | 0 \rangle \nobracket \nobracket & =  - \frac{1}{2} r
  \partial_r \phi (r, 0,0, 0) | 0 \rangle \nobracket \nobracket \,,\\
  L_1 \phi (r, 0,0, 0) | 0 \rangle \nobracket \nobracket & =  0 \,,\\
  M_{1, 1} \phi (r, 0, 0,0) | 0 \rangle \nobracket \nobracket & =  \partial_r
  \phi (r, 0,0, 0) | 0 \rangle \nobracket \nobracket \,,\\
  \left( M_{0, 1} + \frac{1}{r} L_{- 1} \right) \phi (r, 0,0, 0) | 0 \rangle
  \nobracket \nobracket & =  0 \,,
\eag
\ee
and their anti-holomorphic partners.
In order to satisfy the first equation in (\ref{tran1}) and its anti-holomorphic partner,
\be
L_0 \phi (r, 0, 0,0) | 0 \rangle=\bar{L}_0 \phi (r, 0, 0,0) | 0 \rangle\,,
\ee
we have to introduce the term $\left(\frac{\sqrt{2} M_{1, 0}}{m} \right)^{J - L}$ in the construction of our bulk field
\be
\phi (r, 0, 0,0) | 0 \rangle \nobracket \nobracket = \sum^\infty_{J,
L, M=0} \lambda_{J, L, M} r^{- K} L_{- 1}^J \bar{L}_{- 1}^L \left(
\frac{\sqrt{2} M_{1, 0}}{m} \right)^{J - L} M_{0, 0}^M | \mathcal{O}\rangle \nobracket
\nobracket\,.
\ee
The first equation in (\Ref{tran1}) tells
\be
  K =  \Delta + J + L + M\,.
\ee
The second equation in (\Ref{tran1}) and its anti-holomorphic partner tell
\be\label{eq2}
\bag
  \frac{m}{\sqrt{2}} (M + 1) \lambda_{J, L + 1, M + 1} + (L + M + \Delta) (J +
  1) \lambda_{J + 1, L + 1, M} & =  0 \,,\\
  \frac{m}{\sqrt{2}} (M + 1) \lambda_{J + 1, L, M + 1} + (J + M + \Delta) (L +
  1) \lambda_{J + 1, L + 1, M} & =  0 \,.
\eag
\ee
The third equation in (\Ref{tran1}) tells
\be\label{eq3}
  \frac{m}{\sqrt{2}} [(L + 1) \lambda_{J, L + 1, M} + (J + 1) \lambda_{J + 1,
  L, M}] + (L + 1) (J + 1) \lambda_{J + 1, L + 1, M - 1}  =  - K \lambda_{J,
  L, M} \,.
\ee
The fourth equation in (\Ref{tran1}) and its anti-holomorphic partner tell
\be\label{eq4}
\bag
  L \lambda_{J, L, M - 1} + \frac{m}{\sqrt{2}} \lambda_{J, L - 1, M} +
  \lambda_{J - 1, L - 1, M} & =  0 \,,\\
  J \lambda_{J, L, M - 1} + \frac{m}{\sqrt{2}} \lambda_{J - 1, L, M} +
  \lambda_{J - 1, L - 1, M} & =  0 \,.
\eag
\ee
Once these equations are solved, the bulk field at a general point $(r, u, z,\bar{z})$ is obtained by
\be
  \phi (r, u, z,\bar{z}) = \exp (u M_{0, 0}+z L_{-1}+\bar{z}\bar{L}_{-1}) \phi (r, 0, 0) \exp (- u M_{0, 0}-z L_{-1}-\bar{z}\bar{L}_{-1})\,.
\ee
Its action on the vacuum is
\be\label{expan}
  \phi (r, u, z,\bar{z}) | 0 \rangle \nobracket \nobracket 
  =  \sum_{J, L, M} \lambda_{J, L, M} r^{- (\Delta + J + L + M)} e^{z L_{-1}+\bar{z}\bar{L}_{-1}} L_{- 1}^J
  \bar{L}_{- 1}^L \left( \frac{\sqrt{2} M_{1, 0}}{m} \right)^{J - L} M_{0,
  0}^M e^{u M_{0, 0}} | \mathcal{O}\rangle \nobracket \nobracket
\ee

As in the $\tmop{BMS}_3$/$\tmop{CCFT}_2$ case, the Poincar\'e symmetry will be broken unless we take the mass to zero. To show this point, we consider a specific case $\Delta=2$, and find
\be
\lambda_{J,L,M}/\lambda_{0,0,0}=\left(-\frac{\sqrt{2}}{m}\right)^{J+L+2M}\frac{(J+M)!}{J!M!}\frac{(L+M)!}{L!M!}M!\,.
\ee
In this case, the first three equations in (\Ref{tran1}) are satisfied while the fourth one becomes
\be
\left( M_{0, 1} + \frac{1}{r} L_{- 1} \right) \phi (r, 0,0, 0)|0\rangle= \frac{\lambda_{0,0,0}}{r^\Delta}M_{0,1}|\mathcal{O}\rangle\,,
\ee
similar to (\ref{simpcase32}). We will see that the right-hand side becomes zero when considering its massless limit, see (\ref{masslessrep}).

It is remarkable that the newly-proposed highest-weight representation has not been studied in the literature. We do not know the two-point functions\footnote{The two-point functions in CCFTs have been studied from different points of view in \cite{Chen:2021xkw,Chen:2023pqf}. Unfortunately, we cannot apply the results there directly.} in the boundary $\tmop{CCFT}_3$, which strongly prevents us from directly calculating the bulk-boundary propagator as in the $\tmop{BMS}_3$/$\tmop{CCFT}_2$ case.

\subsection{Bulk-boundary propagator in $\tmop{BMS}_4$/$\tmop{CCFT}_3$}\label{section3-3}

Although we do not have the two-point functions of the newly-defined highest-weight representation in the boundary $\tmop{CCFT}_3$, its massless limit is well-understood, which is defined by
\be\label{masslessrep}
M_{1,0}|\mathcal{O}\rangle=M_{0,1}|\mathcal{O}\rangle=0\,.
\ee
Moreover, we can obtain results in the massless limit by considering the bulk-boundary propagator directly, like (\Ref{soph}),
\be
\bag
\langle0|\mathcal{O}(u,z,\bar{z})\phi(r,0,0,0)|0\rangle
=& \left\{|z|^{-2\Delta}\mathcal{O}\left(-\frac{u}{|z|^{2}},\frac{1}{z},\frac{1}{\bar{z}}\right)|0\rangle\right\}^{\dagger}\phi(r,0,0,0)\mid0\rangle   \\
=&\left\{|z|^{-2\Delta} e^{-\frac{u}{|z|^{2}}M_{0,0}+\frac{1}{z}L_{-1}+\frac{1}{\bar{z}}\bar{L}_{-1}}\mathcal{O}\left(0,0,0\right)\mid0\rangle\right\}^{\dagger}\phi(r,0,0,0)\mid0) \\
=&|z|^{-2\Delta}\langle \mathcal{O}|e^{-\frac u{|z|^2}M_{1,1}+\frac{1}{z} L_1+\frac{1}{\bar{z}} \bar{L}_1}\phi(r,0,0,0)|0\rangle\,.  \\
\eag
\ee
Using the second and the third equations in (\ref{tran1}), we read
\be
\bag
|z|^{-2\Delta}\langle \mathcal{O}|e^{-\frac u{|z|^2}M_{1,1}+\frac{1}{z} L_1+\frac{1}{\bar{z}} \bar{L}_1}\phi(r,0,0,0)|0\rangle
=&|z|^{-2\Delta}e^{-\frac{u}{|z|^{2}}\partial_{r}}\langle \mathcal{O}|\phi(r,0,0,0)|0\rangle \\
=&|z|^{-2\Delta}e^{-\frac u{|z|^2}\partial_r}r^{-\Delta}  \\
=&\left(\frac{1}{r|z|^2-u}\right)^\Delta\,.
\eag
\ee
The second equality here, similarly the second equality in (\ref{secequality}), is due to
\be
\bag
\langle \mathcal{O}|\phi(r,0,0,0)|0\rangle&=\sum_{J,L,M}\lambda_{J,L,M} r^{- (\Delta+ J + L + M)}\langle \mathcal{O}|L_{- 1}^J
  \bar{L}_{- 1}^L \left( \frac{\sqrt{2} M_{1, 0}}{m} \right)^{J - L} M_{0,
  0}^M|\mathcal{O}\rangle\\
  &=\sum_{J,L,M}\lambda_{J,L,M} r^{- (\Delta+ J + L + M)}\left[L_{1}^J
  \bar{L}_{1}^L \left( \frac{\sqrt{2} M_{0, 1}}{m} \right)^{J - L} M_{1,
  1}^M|\mathcal{O}\rangle\right]^\dagger |\mathcal{O}\rangle\\
  &=r^{-\Delta}\,.
\eag
\ee
Or equivalently, we have 
\be
\langle\phi(r,u,z,\bar{z})\mathcal{O}(0,0,0)\rangle=\left(\frac{1}{r|z|^2+u}\right)^\Delta\,.
\ee
This is exactly the same as the result using the Kirchhoff-d'Adhemar formula in \cite{Donnay:2022wvx}, which is given by
\be
\bag
\langle\phi(r,u,z,\bar{z})\mathcal{O}(0,0,0)\rangle &=\int d^2 w d \tilde{u} \partial_{\tilde{u}} \delta(\tilde{u}-u-r|z-w|^2)\langle \mathcal{O}(\tilde{u},w,\bar{w})\mathcal{O}(0,0,0)\rangle\\
&=\int d^2 w d \tilde{u} \partial_{\tilde{u}} \delta(\tilde{u}-u-r|z-w|^2) \lim_{\Delta\rightarrow1} \frac{\delta^2(w)}{\tilde{u}^{2\Delta-2}}\\
&\propto\left(\frac{1}{r|z|^2+u}\right)^\Delta\,.
\eag
\ee


\section{Conclusion and discussions}\label{section4}

In this note, we presented a different point of view on bulk reconstruction in flat holography.  We expressed a bulk field as a linear combination of BMS descendants of the highest weight representation, where the operator-state correspondence has already been applied. In contrast, for the induced representations, the operator-state correspondence has not been established. We found that we could not directly reconstruct a massless bulk field due to the weird behavior of the two-point functions of the primary operators of charge $\xi=0$ in CCFT. Our prescription is to start with the reconstruction of a massive bulk field. Even though the reconstruction does not respect the Poincar\'e symmetry, which is in accordance with the fact that a massive particle can never reach the null infinity of flat spacetime, its massless limit is well-behaved. It not only respects the Poincar\'e symmetry but also can be used to derive nontrivial physical quantities like the bulk-boundary propagator. We managed to reproduce the bulk-boundary propagators for the scalars in both BMS$_3$/CCFT$_2$ and BMS$_4$/CCFT$_3$ \cite{Donnay:2022wvx}.

In our study, we considered the highest-weight representation in CCFT. In particular, the ones with $\xi\neq 0$ in $\tmop{CCFT}_3$ have not been studied carefully in the literature. It deserves to have more investigation on their properties.  


In our construction, we only used the global part of the BMS symmetry. In $\tmop{AdS}_3$/$\tmop{CFT}_2$ the Virasoro algebra, which is the asymptotic symmetry for $\tmop{AdS}_3$, can completely control the behavior of the gravity because there is no local physical degree of freedom for the gravity in three dimensions. This could also be the case in $\tmop{BMS}_3$/$\tmop{CCFT}_2$. It is intriguing how to develop the quantum (at least semi-classical) gravity from the full BMS$_3$ symmetry. The situation in $\tmop{BMS}_4$/CCFT$_3$ is more perplexing, as there do exist local physical degrees of freedom in 4d gravity. A natural question is to what extent the $\tmop{BMS}_4$ algebra can control the dynamics of the 4d gravity. We wish our study could shed new light on these issues.




\section*{Acknowledgments}
We thank Pengxiang Hao, Reiko Liu, Haowei Sun and Yufan Zheng very much for their valuable discussions. This work is in part supported by NSFC Grant  No. 12275004, 11735001.

\bibliographystyle{utphys}
\bibliography{refs}
\end{document}